\documentclass[preprint,preprintnumbers,showkeys]{revtex4}

\begin{document}

\title{Quark Confinement and the Renormalization Group}

\author{Michael C. Ogilvie}

\affiliation{Dept. of Physics, Washington University, St. Louis, MO 63130 USA}

\label{firstpage}


% keywords for Phil Trans commented out
\begin{abstract}%{renormalization group, qcd, quark confinement}
Recent approaches to quark confinement are reviewed, with an emphasis
on their connection to renormalization group methods. Basic concepts
related to confinement are introduced: the string tension, Wilson
loops and Polyakov lines, string breaking, string tension scaling
laws, center symmetry breaking, and the deconfinement transition at
non-zero temperature. Current topics discussed include confinement
on $R^{3}\times S^{1}$, the real-space renormalization group, the
functional renormalization group, and the Schwinger-Dyson equation
approach to confinement.
\end{abstract}

\preprint{INT-PUB-10-046}

\maketitle

\section{Introduction}

Quarks are the fermionic constituents of the hadrons, the strongly
interacting particles such as the proton and the neutron. The problem
of quark confinement, at its simplest, is to explain why quarks are
not observed as physical states in high-energy scattering, but remain
confined inside hadrons. The problem has been apparent for decades,
and pre-dates quantum chromodynamics (QCD), the modern theory of the
strong interactions \cite{Greenberg:1964pe,Politzer:1973fx,Gross:1973ju}. 
Lattice gauge theory \cite{Wilson:1974sk,Creutz:1980zw} has given us a physical
basis for quark confinement: the force between quarks is independent
of distance at large separation, and an infinite amount of energy
would be required to isolate a single quark. The force constant at
infinite separation is the string tension $\sigma$. While lattice
simulations have determined the relation of $\sigma$ to other observables
of QCD with high precision, we are still far from a satisfactory understanding
of quark confinement. We have neither a firm grasp of the mechanisms
of a quark confinement nor the ability to calculate analytically even
simple aspects of the physics of confinement.

Recent work in the study of confinement suggests that progress is
now being made, and there are several complementary approaches
under development. All of these approaches have common features: they
use the renormalization group in an essential way, and are strongly
conditioned by lattice gauge theory results. This review is based
on talks given at the INT Workshop "New applications of
the renormalization group in nuclear, particle,
and condensed matter physics", 
held February 22-26 2010,
and the emphasis is on current research
rather than a comprehensive treatment
of each approach.
The review is organized as follows:
Section II introduces the basic concepts involved in the study of
quark confinement: the string tension, its measurement, string breaking,
the finite-temperature deconfinement transition, center symmetry,
and string tension scaling laws. Section III discusses recent progress
in understanding confinement on $R^{3}\times S^{1}$ Section IV describes
recent work on a rigorous proof of confinement using real-space renormalization
group methods. Section V discusses two related approaches to confinement
based on the infrared behavior of quark and gluon propagators, the
first using the functional renormalization group and the second using
the Schwinger-Dyson equations. A final section attempts to summarize
our current understanding and prospects for future progress.

\section{Fundamentals of confinement}

The fundamental theory of the strong interactions is quantum chromodynamics
(QCD), a quantum field theory describing the interactions of quarks and gluons.
QCD is a gauge theory: Gluons are massless spin-1 bosons whose interactions
are tightly constrained by the requirement of local gauge invariance in a manner
similar to the case of QED (quantum electrodynamics) and photons.
In QED, the symmetry group is $U(1)$, the group of local phase transformations
on charged fields: $\phi ( x ) \rightarrow e^{i\alpha (x)} \phi ( x )$.
In QCD, the corresponding symmetry group is $SU(3)$, and the symmetry
acts on the three color charges carried by quarks, 
usually labeled as red, blue and green.
There is a profound difference between QED and QCD: the gauge group
$U(1)$ of QED is Abelian, but the gauge group $SU(3)$ of QCD
is non-Abelian. Physically, this implies that the photon has no electric charge
but gluons do carry color charge. 
The four-vector field of the photon is written as $A_\mu (x)$ where
$\mu=1,2,3,4$, but an $SU(N)$ gauge field is written as $A_\mu^a (x)$,
where the index $a$ runs over the $N^2-1$ one charges of the
non-Abelian gauge field.
While the photon is
a free field with a trivial renormalization group beta function,
$SU(3)$ gauge theory, even without quarks, is a theory
with highly non-trivial interactions. The beta function
of an $SU(N)$ gauge theory without quarks
is
\begin{equation}
\beta(g) = - {g^3 \over (4\pi)^2} {11 N \over 3 }
\end{equation}
to lowest order in perturbation theory.
Here $g$ is the strong-interaction coupling constant.
It is actually a running coupling constant $g(\mu )$,
depending on the momentum-space renormalization
scale $\mu$, evolving according to
\begin{equation}
\mu { dg \over d \mu } = \beta \left( g\left( \mu \right) \right).
\end{equation}
The negative sign of $\beta\left( g \right)$ indicates
that $SU(N)$ gauge theories are asymptotically free:
There is an ultraviolet fixed point at $g=0$, and the
interactions become weak at high energies.
There is a corresponding infrared fixed point
at $g=\infty$, indicating that
the running coupling $g(\mu )$ becomes large
at low energies. This implies that the low-energy
properties of QCD, unlike QED, cannot be analyzed
via perturbation theory. 
The most effective tool we have
for exploring the non-perturbative
aspects of QCD is lattice gauge theory,
in which continuous space-time
is replaced by discrete lattice,
with a fundamental cut-off scale
set by the lattice spacing $a$.
The renormalization group
plays a crucial role in the
taking of the continuum limit,
where $a\rightarrow 0$.
In gauge theories, 
the gauge field has four-components 
but two physical polarization states. 
Two
of the four components of the field correspond to redundant degrees
of freedom, and must be eliminated by a choice of gauge, such as
Coulomb gauge or Landau gauge, for continuum
field theory methods to be applied.
Lattice gauge theory, on the other hand,
maintains exact gauge invariance:
no gauge fixing is required, although
it may be applied to lattice gauge theories.
Physical predictions of theory, such
as hadron masses, must of course
be independent of the choice of gauge.
A great deal of the technical problems
in analytic treatments of non-perturbative
aspects of gauge theories are
ultimately associated with achieving
gauge invariance and 
correct renormalization
group scaling behavior simultaneously.

%\subsubsection*{String tension via the Wilson loop}

The static potential between two heavy quarks can be determined in
a gauge-invariant way from the expectation value of the Wilson loop.
This is a non-local operator associated with a closed curve $\mathcal{C}$
in space-time parametrized as $x_{\mu}\left(\tau\right)$ where $\tau$
can be taken to be the unit interval and $x_{\mu}\left(1\right)=x_{\mu}\left(0\right)$.
The Wilson loop is defined as\begin{equation}
W\left[\mathcal{C}\right]=\mathcal{P}\exp\left[i\oint_{\mathcal{C}}dx_{\mu}A_{\mu}\left(x\right)\right]\end{equation}
where $\mathcal{P}$ indicates path-ordering of the gauge fields 
$A_\mu\left(x\right)$ along
the path $\mathcal{C}$. Taking $W\left[\mathcal{C}\right]$ to be
an element of the abstract gauge group $G$, the basic observables
are $Tr_{R}W\left[\mathcal{C}\right]$, where the trace is taken over
an irreducible representation $R$ of $G$. $Tr_{R}W\left[\mathcal{C}\right]$
has a physical interpretation of the non-Abelian phase factor associated
with a heavy particle in the representation $R$ moving adiabatically
around the closed loop $\mathcal{C}$. Typically, we are interested
in rectangular Wilson loops of with sides $L$ and $T$. The loop
can be associated with a process in which a particle-antiparticle
pair are created at one time, move to a separation $L$, propagate
forward in time for an interval $T$, and then annihilate. For $T\gg L$,
we have\begin{equation}
\left< Tr_{R}W\left[\mathcal{C}\right] \right>
\simeq\exp\left[-V_{R}\left(L\right)T\right]\end{equation}
where $V_{R}\left(L\right)$ is the heavy quark-antiquark potential
for particles in the representation $R$. For large distances, the
potential can grow no faster than linearly with $L$, and that linear
growth defines the string tension $\sigma_{R}$ for the representation
$R$:\begin{equation}
V\left(L\right)\rightarrow\sigma_{R}L+\mathcal{O}\left(1\right)\end{equation}
 as $L\rightarrow\infty$. The $\mathcal{O}\left(1\right)$ correction
term represents a so-called perimeter law contribution, which is not
physical; there are also $1/L$ corrections which are of physical
interest. The statement that quarks are confined is the statement
that the string tension $\sigma_{F}$ in the fundamental representation
of $SU(3)$ gauge theory is non-zero. A characteristic feature of
pure gauge theories in four dimensions is dimensional transmutation:
the classical action is scale-invariant, but the introduction of a
mass scale $\mu$ is necessary in order to define the running coupling
constant $g(\mu)$. It follows directly from the renormalization group
that any physical mass in four-dimensional pure gauge theories must
be proportional to a renormalization group invariant mass, usually
written simply as $\Lambda$. This implies that $\sigma_{R}=c_{R}\Lambda^{2}$,
where $c_{R}$ is a pure number.

%\subsubsection*{Role of quarks and string breaking}

Paradoxically, string tensions are generally determined in so-called
pure gauge theories, where only the gauge fields are dynamical; quarks
and particles exist only as static classical charges. This is done
because of string breaking , also known as charge screening. Consider
a theory with dynamical particles of mass $m$ in the fundamental
representation $F$ of the gauge group. Then energy obtained by separating
a pair of static sources in the fundmental representation is $\sigma_{F}L$;
when that energy becomes on the order of $2m$, it will become energetically
favorable to produce a pair of dynamical particles at an energy cost
$2m$, and no string tension will be seen for $L\gtrsim2m/\sigma_{F}$.

%\subsubsection*{String tension via Polyakov loop}

If one or more directions in space-time are compact, the string tension
may also be determined using the Polyakov loop $P$, also known as
the Wilson line. The Polyakov loop is essentially a Wilson loop that
uses a compact direction in space-time to close the curve using a
topologically non-trivial path in space time. The typical use for
the Polyakov loop is for gauge theories at finite temperature, where
space-time is $R^{3}\times S^{1}$. The partition function being given
by $Z=Tr\left[e^{-\beta H}\right]$, the circumference of $S^{1}$
is given by the inverse temperature $\beta=1/T$. In this case, we
write\begin{equation}
P\left(\vec{x}\right)=\mathcal{P}\exp\left[i\int_{0}^{\beta}dx_{4}A_{4}\left(x\right)\right]\end{equation}
 and string tensions may be determined from a two-point function\begin{equation}
\left\langle Tr_{R}P\left(\vec{x}\right)Tr_{R}P^{\dagger}\left(\vec{y}\right)\right\rangle \sim e^{-\sigma_{R}\left|\vec{x}-\vec{y}\right|}\end{equation}
 a behavior that assumes that the one-point function $\left\langle Tr_{R}P\left(\vec{x}\right)\right\rangle =0$.
The Polyakov loop one-point function $\left\langle Tr_{R}P\left(\vec{x}\right)\right\rangle $
can be interpreted as the Boltzman factor $\exp\left(-\beta F_{R}\right)$,
where $F_{R}$ is the free energy required to add a static particle
in the representation $R$ to the system. Of course, $\left\langle Tr_{R}P\left(\vec{x}\right)\right\rangle =0$
implies that $F_{R}=\infty$ , which is thus a fundamental criterion
determining whether particles in the representation $R$ are confined.

%\subsubsection*{Z(N) symmetry}

One of the most important concepts in our understanding of confinement
is the role of center symmetry. The center of a Lie group is the set
of all elements that commute with every other element. For $SU(N)$,
this is $Z(N)$. Although the $Z(N)$ symmetry of $SU(N)$ gauge theories
can be understood from the continuum theory, it is easier to understand
from a lattice point of view. A lattice gauge theory associates link
variable $U_{\mu}\left(x\right)$ with each lattice site $x$ and
direction $\mu$. The link variable is considered to be the path-ordered
exponential of the gauge field from $x$ to $x+\hat{\mu}$:
$U_{\mu}\left(x\right) = \exp\left[ i A_\mu\left(x\right)\right].$
Consider
a center symmetry transformation on all the links in a given direction
on a fixed hyperplane perpendicular to the direction. The standard
example from $SU(N)$ gauge theories at finite temperature is $U_{4}\left(\vec{x},t\right)\rightarrow zU_{4}\left(\vec{x},t\right)$
for all $\vec{x}$ and fixed $t$, with $z\in Z(N)$. Because lattice
actions such as the Wilson action consist of sums of small Wilson
loops, they are invariant under this global symmetry. However, the
Polyakov loop transforms as $P\left(\vec{x}\right)\rightarrow zP\left(\vec{x}\right)$,
and more generally\begin{equation}
Tr_{R}P\left(\vec{x}\right)\rightarrow z^{k_{R}}Tr_{R}P\left(\vec{x}\right)\end{equation}
 where $k_{R}$ is an integer in the set $\left\{ 0,1,...,N-1\right\} $
and is known as the $N$-ality of the representation $R$. If $k_{R}\ne0$,
then unbroken global $Z(N)$ symmetry implies $\left\langle Tr_{R}P\left(\vec{x}\right)\right\rangle =0$.
Thus global $Z(N)$ symmetry defines the confining phase of a gauge
theory. For pure gauge theories at non-zero temperature, the deconfinment
phase transition is associated with the loss of $Z(N)$ symmetry at
the critical point $T_{d}$. Below that point $\left\langle Tr_{F}P\left(\vec{x}\right)\right\rangle =0$
but above $T_{d}$, $\left\langle Tr_{R}P\left(\vec{x}\right)\right\rangle \ne0$.

Notice that the case of zero $N$-ality representations is special
within this framework: there is no requirement from $Z(N)$ symmetry
that these representations are confined. This includes the adjoint
representation, the representation of the gauge particles. However,
lattice simulation indicate that $\left\langle Tr_{R}P\left(\vec{x}\right)\right\rangle $
is very small for these representations in the confined phase. Although
screening by gauge particles must dominate at large distances, these
zero $N$-ality representations have well-defined string tensions
at intermediate distances scales, $e.g.$, on the order of a few fermi
for $SU(3)$, behaving in a manner very similar to representations
with non-zero $N$-ality \cite{Deldar:1999vi,Bali:2000un}.

%\subsubsection*{Different string tensions}

It is known from lattice simulations that each representation apparently
has its own distinct string tension at intermediate distance scales.
To a good approximation, the scaling behavior observed in lattice
simulations is consistent with Casimir scaling\begin{equation}
\frac{\sigma_{R}}{\sigma_{F}}=\frac{C_{R}}{C_{F}}\end{equation}
where $C_{R}$ is the quadratic Casimir operator for the representation
$R$. This behavior is exact in two-dimensional pure gauge theories.
It is generally believed that at sufficiently large distance scales,
screening by gauge particles will lead to a single string tension
for all representations within a given $N$-ality class, \emph{i.e.},
with the same value of $k$. Labeling the lightest value of $\sigma_{R}$
within an $N$-ality class by $\sigma_{k}$, Casimir scaling predicts\begin{equation}
\frac{\sigma_{k}}{\sigma_{1}}=\frac{k\left(N-k\right)}{N-1}\end{equation}
 However, the observed behavior is also consistent with so-called
sine-Law scaling\begin{equation}
\frac{\sigma_{k}}{\sigma_{1}}=\frac{\sin\left(\frac{\pi k}{N}\right)}{\sin\left(\frac{\pi}{N}\right)}\end{equation}
 and lattice results have not yet been definitive \cite{Greensite:2003bk}.

\section{Confinement on $R^{3}\times S^{1}$}

In the last few years, it has proven possible to construct four-dimensional
gauge theories for which confinement may be reliably demonstrated
using semiclassical methods \cite{Myers:2007vc,Unsal:2007vu}. 
These models combine $Z(N)$ symmetry,
the effective potential for $P$, instantons, and monopoles into a satisfying picture
of confinement for a special class of models. All of the models in
this class have one or more small compact directions. Models with
an $R^{3}\times S^{1}$ topology have been most investigated, and
discussion here will be restricted to this class. For recent work
in other compactifications, see \cite{Azeyanagi:2010ne}. The use of one or more compact
directions will cause the running coupling constant of an asymptotically
free gauge theory to be small, so that semiclassical methods are reliable.
For example, if the circumference $L$ of $S^{1}$ is small, \emph{i.e.},
$L\ll\Lambda^{-1}$ , then $g(L)\ll1$. However, this leads to an
immediate problem: generally speaking, one or more small compact directions
lead to breaking of $Z(N)$ symmetry in those directions. This has
been clear for many years for the case of finite temperature gauge
theories, where $L=\beta=1/T$. The effective potential for
the Polyakov loop is easily calculated to lowest order
in perturbation theory; for a pure gauge theory it
is given by\begin{equation}
V_{gauge}\left(P,\beta\right)=\frac{-2}{\pi^{2}\beta^{4}}\sum_{n=1}^{\infty}\frac{Tr_{A}P^{n}}{n^{4}}\end{equation}
where the trace of $P$ in the adjoint representation is given by
$Tr_{A}P=\left|Tr_{F}P\right|^{2}-1$. This effective potential is
minimized when $P=zI$ where $z\in Z(N)$, indicating that $Z(N)$
symmetry is spontaneously broken at high temperatures where the one-loop
expression is valid. Note that when $P=zI$, the effective potential
is just the free energy density of a blackbody with $2\left(N^{2}-1\right)$
degrees of freedom. It can easily be shown that any additional fields
also act to break $Z(N)$ symmetry at high temperatures, corresponding
to small $L=\beta$.

There are two broad approaches to maintaining $Z(N)$ symmetry for
small $L$. The first approach deforms the pure gauge theory by adding
additional terms to the gauge action \cite{Myers:2007vc,Ogilvie:2007tj,Unsal:2008ch}. 
The general form for such a
deformation is\begin{equation}
S\rightarrow S+\beta\int d^{3}x\,\sum_{k=1}^{\infty}a_{k}Tr_{A}P\left(\vec{x}\right)^{k}.
\end{equation}
Such terms are often referred to as double-trace deformations.
 If the coefficients $a_{k}$ are sufficiently large, they will counteract
the effects of the one-loop effective potential, and $Z(N)$ symmetry
will hold for small $L$. Strictly speaking, only the first $\left[N/2\right]$
terms are necessary to ensure confinement. It is easy to prove that
for a classical Polyakov loop $P$, the conditions $Tr_{F}P^{k}=0$
with $1\le k\le\left[N/2\right]$ determine the unique set of Polyakov
loop eigenvalues that constitute a confining solution, \emph{i.e.},
one for which $Tr_{R}P=0$ 
for all representations with $k_{R}\ne0$ \cite{Meisinger:2001cq}.
The explicit solution is interesting: up to a factor necessary to
ensure $\det P$=1, the eigenvalues of $P$ are given by the set of $N$'th roots of unity,
which are permuted by a global $Z(N)$ symmetry transformation. The
effective potential associated with $S$ is given approximately by\begin{equation}
V_{1-loop}\left(P,\beta\right)=\frac{-2}{\pi^{2}\beta^{4}}\sum_{n=1}^{\infty}\frac{Tr_{A}P^{n}}{n^{4}}+\sum_{k=1}^{\left[\frac{N}{2}\right]}a_{k}Tr_{A}P^{k}.\end{equation}
For sufficiently large and positive values of the $a_k$'s,
the confined phase yields the lowest value of $V_{1-loop}$.
However, a rich phase structure emerges from the
minimization of $V_{1-loop}$. 
For $N\ge3$, the effective potential predicts that
one or more phases separate the deconfined phase from the confined
phase. In the case of $SU(3)$, a single new phase is predicted, and
has been observed in lattice simulations \cite{Myers:2007vc}. 
For larger values of $N$,
there is a rich set of possible phases, including some where $Z(N)$
breaks down to a proper subgroup $Z(p)$. In such phases, particles
in the fundamental representation are confined, but bound states of
$N/p$ such particles are not 
\cite{Ogilvie:2007tj}.

Lattice simulations of $SU(3)$ and $SU(4)$ agree with the theoretical
predictions based on effective potential arguments \cite{Myers:2007vc}. 
The phase diagram
of $SU(3)$ as a function of $T=L^{-1}$ and $a_{1}$ has three phases:
the confined phase, the deconfined phase, and another phase, the skewed
phase, where charge-conjugation symmetry is spontaneously broken.
In the case of $SU(4)$, a sufficiently large value of $a_{1}$ leads
to a partially-confining phase where $Z(4)$ is spontaneously broken
to $Z(2)$. Particles with $k=1$ are confined in this phase, \emph{i.e.},
$ $$\left\langle Tr_{F}P\left(\vec{x}\right)\right\rangle =0$, but
particles with $k=2$ are not, as indicated by $\left\langle Tr_{F}P^{2}\left(\vec{x}\right)\right\rangle \ne0$.
An important result obtained from the lattice simulation of $SU(3)$
is that the small-$L$ confining region, where semiclassical methods
yield confinement, are smoothly connected to the conventional large-$L$
confining region.

%\subsubsection*{Adjoint fermions}

Another approach to preserving $Z(N)$ symmetry for small $L$ uses
fermions in the adjoint representation with periodic boundary conditions
in the compact direction \cite{Unsal:2007vu}. 
In this case, it is somewhat misleading
to use $\beta$ as a synonym for $L$, because the transfer
matrix for evolution in the compact direction is not positive-definite.
Periodic boundary conditions in the timelike direction imply that
the generating function of the ensemble, \emph{i.e.}, the partition
function, is given by\begin{equation}
Z=Tr\left[\left(-1\right)^{F}e^{-\beta H}\right]\end{equation}
 where $F$ is the fermion number. This graded ensemble, familiar
from supersymmetry, can be obtained from an ensemble $Tr\left[\exp\left(\beta\mu F-\beta H\right)\right]$
with chemical potential $\mu$ by the replacement $\beta\mu\rightarrow i\pi$.
This system can be viewed as a gauge theory with periodic boundary
conditions in one compact spatial direction of length $L=\beta$,
and the transfer matrix in the time direction is positive-definite,

The use of periodic boundary conditions for the adjoint fermions dramatically
changes their contribution to the Polyakov loop effective potential.
In perturbation theory, the replacement $\beta\mu\rightarrow i\pi$
shifts the Matsubara frequencies from $\beta\omega_{n}=\left(2n+1\right)\pi$
to $\beta\omega_{n}=2n\pi$. The one loop effective potential is now
essentially that of a bosonic field, but with an overall negative sign due to
fermi statistics 
\cite{Meisinger:2001fi}.
The sum
of the effective potential for the fermions plus that of the gauge
bosons gives \begin{equation}
V_{1-loop}\left(P,\beta,m,N_{f}\right)=\frac{1}{\pi^{2}\beta^{4}}\sum_{n=1}^{\infty}\frac{Tr_{A}P^{n}}{n^{2}}\left[2N_{f}\beta^{2}m^{2}K_{2}\left(n\beta m\right)-\frac{2}{n^{2}}\right]\end{equation}
where $N_{f}$ is the number of adjoint Dirac fermions and $m$ is
their mass. Note that the first term in brackets, due to the fermions,
is positive for every value of $n$, while the second term, due to
the gauge bosons, is negative. 

The largest contribution to the effective potential at high temperatures
is typically from the $n=1$ term, which can be written simply as
\begin{equation}
\frac{1}{\pi^{2}\beta^{4}}\left[2N_{f}\beta^{2}m^{2}K_{2}\left(\beta m\right)-2\right]\left[\left|Tr_{F}P\right|^{2}-1\right]\end{equation}
 where the overall sign depends only on $N_{f}$ and $\beta m$ \cite{Meisinger:2009ne}. 
 If $N_{f}\ge1$ and $\beta m$ is sufficiently small, this term will
favor $Tr_{F}P=0$. On the other hand, if $\beta m$ is sufficiently
large, a value of $P$ from the center, $Z(N)$, is preferred. Note
that an $\mathcal{N}=1$ super Yang-Mills theory would correspond
to $N_{f}=1/2$ and $m=0,$ giving a vanishing perturbative contribution
for all $n$ 
\cite{Davies:1999uw,Davies:2000nw}.
In that case, non-perturbative effects lead to a confining effective
potential for all values of $\beta$. In the case of $N_{f}\ge1$,
each term in the effective potential will change sign in succession
as $m$ is lowered towards zero. For larger values of $N$, this leads
to a cascade of phases separating the confined and deconfined phases.
%Numerical investigation shows that the confined phase is obtained
%if $N\beta m\lesssim XXXXX$. 
As $m$ increases, it becomes favorable
that $Tr_{F}P^{n}\ne0$ for successive values of $n$. If $N$ is
even, the first phase after the confined phase will be a phase with
$Z(N/2)$ symmetry. As $m$ increases, the last phase before reaching
the deconfined phase will have $Z(2)$ symmetry, in which $k=1$ states
are confined, but all states with higher $k$ are not.
Lattice simulations of $SU(3)$ with
periodic adjoint fermions are completely
consistent with this picture\cite{Cossu:2009sq}.
For $N\ge 3$, 
there are generally phases intermediate between the confined
and deconfined phases which are not of the partially-confined type.
Careful numerical analysis appears to be necessary on a case-by-case
basis to determine the phase structure for each value of $N$
 \cite{Myers:2009df}.

There are some interesting additional issues arising when periodic
adjoint fermions are used to obtain $Z(N)$ symmetry for small $L$.
There are strong indications from strong-coupling lattice calculations
 \cite{Gocksch:1984yk}
and the closely related Polyakov-Nambu-Jona Lasinio (PNJL) models 
\cite{Fukushima:2003fw}
that the
mass $m$ that appears in the effective potential $V(P)$ should be
regarded as a constituent mass that includes the substantial effects
of chiral symmetry breaking. In strong-coupling lattice calculations
and PNJL models, this effect is responsible for the coupling of $P$
and $\bar{\psi}\psi$ in simulations of QCD at finite temperature.
Thus it is important that recent simulations of $SU(3)$ with $N_{f}=2$
flavors of adjoint fermions show explicitly that the $Z(N)$-invariant
confined phase is regained when the fermion mass is sufficiently small.
In fact, the simulations show the behavior expected on the basis of
effective potential arguments: a skewed phase separates the confined
phase and deconfined phase. However, this raises another issue: For
a simple double-trace deformation, the small-$L$ and large-$L$ regions
are smoothly connected. Is this also the case if periodic adjoint
fermions are used? A semi-phenomenological analysis based on a PNJL
model
\cite{Nishimura:2009me}
suggests that the answer is yes. However, there
may be problems in reaching sufficienly light fermion masses in lattice
simulations to confirm this.

%\subsubsection*{Timelike string tensions}

In addition to the overall phase structure, semiclassical methods
may also be used to determine string tensions for small $L$;
the same methods are used with double-trace deformations
and with periodic adjoint fermions.
There are two classes of string tensions, depending on whether the Polyakov loop or the Wilson
loop is used. For simplicity, we will refer to the former case as
electric and the latter as magnetic, borrowing the terminology from
finite temperature gauge theory.
% PARAGRAPH
Electric string tensions are calculable perturbatively in the high-temperature
confining region from small fluctuations about the confining minimum
of the effective potential \cite{Meisinger:2004pa,Meisinger:2009ne}.
%=========== SHORTENED ===================
The $m=0$ limit has the simple form 
\begin{equation}\label{eq:spatialtension}
\left(\sigma_{k}^{(t)} L\right)^{2}=\frac{\left(2N_{f}-1\right)g^{2}T^{2}}{3N}\left[3\csc^{2}\left(\frac{\pi k}{N}\right)-1\right]\end{equation}
and is a good approximation for $\beta m\ll1$.
The confining minimum of the effective potential breaks $SU(N)$ to
$U(1)^{N-1}$. This remaining unbroken Abelian gauge group naively
seems to preclude spatial confinement, in the sense of area law behavior
for spatial Wilson loops. However, as first discussed by Polyakov
in the case of an $SU(2)$ Higgs model in $(2+1)$-dimensional gauge
systems, instantons can lead to nonperturbative confinement 
\cite{Polyakov:1976fu}.
%\textbackslash{}cite\{Polyakov:1976fu\}.
% LONG VERSION
%In the high-temperature confining region, the dynamics of the magnetic
%sector are effectively three-dimensional due to dimensional reduction.
%The Polyakov loop plays a role similar to an adjoint Higgs field,
%with the important difference that $P$ lies in the gauge group, while
%a Higgs field would lie in the gauge algebra. The standard topological
%analysis \textbackslash{}cite\{Weinberg:1979zt\} is therefore slightly
%altered, and there are $N$ fundamental monopoles in the finite temperature
%gauge theory. This behavior was first seen in the construction of
%finite temperature instantons (calorons) with non-trivial Polyakov
%loop behavior at infinity. \textbackslash{}cite\{Lee:1998vu,Kraan:1998kp,Lee:1998bb,Kraan:1998pm,Kraan:1998sn\}
%Each instanton solution contains all $N$ fundamental monopoles. 
There are $N$  different fundamental monopoles in the
gauge theory on $R^3 \times S^1$. 
This behavior was first seen in the construction of
finite temperature instantons (calorons) with non-trivial Polyakov
loop behavior at infinity \cite{Lee:1998vu,Kraan:1998kp,Lee:1998bb,Kraan:1998pm,Kraan:1998sn}.
% ========= SHORT VERSION ===========
As in the case of Polyakov's original model,
the string tension is proportional to
$\exp\left( - S_{monopole} \right)$,
where $S_{monopole}$ is the action of a single monopole.
The other factors are essentially determined by the renormalization
group up to an overall numerical factor \cite{Zarembo:1995am,Unsal:2008ch}.

There remain, however, a number of interesting
issues which require further analytic study, including string
tension scaling laws for Wilson loops, the mechanism for chiral symmetry
breaking, the approach to the large-$N$ limit of $SU(N)$, geometries
with more than one compact direction \cite{Azeyanagi:2010ne}, and gauge groups other than
$SU(N)$. Lattice simulations have the potential to provide much useful
information, for example on the connection between confinement at
small $L$ and large $L$. Because of the small value of the running
coupling constant at small $L$, gauge theories on $R^{3}\times S^{1}$
are a natural area where lattice simulations and analytic calculation
can work together effectively.

\section{Confinement via real-space renormalization group}

$SU(N)$ gauge theories in four dimensions have a single, Gaussian
ultra-violet fixed point at $g^{2}=0$. This means that QCD with massless
fermions has no free parameters once a renormalization-group invariant
length scale $\Lambda$ is defined. However, non-perturbative features
can so far only be determined in lattice simulations. The natural
framework for dealing with such problems is a Wilsonian RG blocking
procedure bridging the different scale regimes. A variety of lattice
gauge theory approaches have been developed, including techniques
for finding the action along the Wilsonian renormalization group trajectory
({}``perfect action'') and Monte Carlo renormalization group techniques.
An alternative approach which has been extensively developed by Tomboulis
is based on approximate real-space renormalization group transformations
which satisfy rigorous bounds 
\cite{Tomboulis:2007iw,Tomboulis:2007sq,Tomboulis:2008zz,Tomboulis:2009zz}. 
The general strategy is to employ
approximate but easily explicitly computable renormalization group
transformations that provide bounds on the ratios of partition functions,
leading to results on free energy differences. Sufficiently strong
bounds constrain will constrain the exact free energy differences.

The Migdal-Kadanoff approximation 
provides a starting point for rigorous analysis
 \cite{Migdal:1975zf,Migdal:1975zg,Kadanoff:1976jb,Kadanoff:1976wy}.
The Migdal-Kadanoff procedure uses a combination of 
decimation and bond-moving
to construct an approximate renormalization group which
gives a rigorous lower bound on free energies. 
Decimation refers to the procedure of integrating over
some of the variable of a lattice system, {\it e.g.}
a lattice spin system such as the Ising model or
a lattice quantum field theory.
Suppose the partition function for such a system
is defined on lattice $\Lambda$ as
$Z_\Lambda = Tr_\Lambda \exp (-S)$
where $S$ is the Euclidean action
for a field theory or $\beta H$ for a classical system.
The $Tr_\Lambda$ indicates that all variables
on the lattice $\Lambda$ are to be integrated over.
One way to construct a real-space renormalization
group transformation is to integrate over some of
the variables, leaving only the variables on a
sub-lattice $\Lambda'$ and a new action $S'$
for those variables. This may be written as
\begin{equation}
\exp ( - S' ) = Tr_{\Lambda/\Lambda'} \exp (-S)
\end{equation}
where the trace is over variables on the complement
$\Lambda/\Lambda'$ of $\Lambda$.
A typical choice for $\Lambda'$ is to go
from a lattice with lattice spacing $a$
to a sub-lattice of the same type with
lattice spacing $b a$ with $b > 1$.
However, this procedure can only be carried
analytically for the simplest of models,
notably $d=1$ spin systems and $d=2$ gauge
theories with local interactions.
In higher dimensions, new, complicated
quasi-local interactions are generated.
The Migdal-Kadanoff procedure uses bond-moving
to restrict the terms generated by decimation.
Bond moving refers
to modifying the lattice action from $S$ to
$S+ \Delta S$ in such a way that the average
action is unchanged: $ Tr_\Lambda \left[ e^S \Delta S \right] = 0$.
In the simplest case, $\Delta S$ decouples
some of the variables on $\Lambda/\Lambda'$
from those on $\Lambda'$ to eliminate unwanted couplings.
This procedure gives a bound ${Z'}_{\Lambda'}$ on the original partition function
\begin{eqnarray}
{Z'}_{\Lambda'}&=& Tr_{\Lambda} \exp (-S-\Delta S)
=Z_\Lambda \left( {{Z'}_{\Lambda'} \over Z_\Lambda}\right) \ge Z_\Lambda
\end{eqnarray}
where Jensen's inequality has been
used in the form 
$\left< e^{-\Delta S} \right>_\Lambda \ge e^{\left< -\Delta S \right>_\Lambda }=1$.
With a good choice for $\Delta S$, this procedure
leads to a relatively simple set of renormalization group equations
for the coupling constants of the model.
This procedure can be generalized in principle
to include blocking of variables, in which
a variable on $\Lambda'$ represents a weighted
average of variables on $\Lambda$.
The Migdal-Kadanoff approach has been usefully
applied in lattice gauge theory to give an understanding of 
renormalization group flows
\cite{Bitar:1982bp,Bitar:1984aj}
and the deconfinement transition \cite{Ogilvie:1983ss,Imachi:1987ps}
in lattice gauge theories.

Tomboulis has generalized the Migdal-Kadanoff procedure
in such a way that not only is the partition
function left invariant, but other quantities
are left invariant as well.
Suppose the goal is to calculate or estimate an expectation value\begin{equation}
\left\langle O\right\rangle =\frac{Z_{\Lambda}\left[O\right]}{Z_{\Lambda}}\end{equation}
The goal is then to construct a renormalization group procedure consisting
of a set of steps\begin{eqnarray*}
a & \rightarrow & ba\rightarrow b^{2}a\rightarrow...\rightarrow b^{n}a\\
\Lambda & \rightarrow & \Lambda^{(1)}\rightarrow\Lambda^{(2)}\rightarrow...\rightarrow\Lambda^{(n)}\end{eqnarray*}
such that
\begin{equation}
\left\langle O\right\rangle =\frac{Z_{\Lambda}\left[O\right]}{Z_{\Lambda}}=...=\frac{Z_{\Lambda_{m}}\left[O\right]}{Z_{\Lambda_{m}}}=...=\frac{Z_{\Lambda_{n}}\left[O\right]}{Z_{\Lambda_{n}}}\end{equation}
After a sufficiently large number of renormalization group transformations,
an asymptotically free theory will reach the strong-coupling region
near the infra-red fixed point at $g^{2}=\infty$, where cluster expansion
methods can be used reliably. Note that the construction is specific
to a given operator $O$: different operators will give rise to different
renormalization group flows. In principle, the difference between
the two flows would be irrelevant in the continuum limit.

The prototypical application of this approach is to the twisted partition
function of an $SU(2)$ lattice gauge theory \cite{Tomboulis:1982px} . 
The twisted partition
function $Z_{\Lambda}^{(-)}$ on an $L_{1}\times L_{2}\times L_{3}\times L_{4}$
lattice $\Lambda$ is obtained from the usual partition function $Z_{\Lambda}$
by modifying the action such that plaquettes in the $x-y$ plane with
fixed values $x_{0}$ and $y_{0}$ of $x$ and $y$ are multiplied
in the action by $-1$ for all values of $z$ and $t$: $U_{xy}\left(x_{0},y_{0},z,t\right)\rightarrow-U_{xy}\left(x_{0},y_{0},z,t\right)$.
The twisted partition function was originally defined in the continuum
by 't Hooft and plays a central role in his analysis of the allowed
phases of gauge theories \cite{'tHooft:1977hy,'tHooft:1979uj}. 
The vortex free energy $F_{\Lambda}$
is defined by the ratio of the two partition functions: $\exp\left[-F_{\Lambda}\right]=Z_{\Lambda}^{(-)}/Z_{\Lambda}$;
in this ratio the bulk free energy term proportional to space-time
volume cancels. 't Hooft's analysis shows that confinement holds if\begin{equation}
F_{\Lambda}\sim cL_{3}L_{4}e^{-\rho L_{1}L_{2}}\end{equation}
with $\rho>0$ as the infinite space-time volume limit is taken with
$L_{1}L_{2}\gg\log\left(L_{3}L_{4}\right)$. For lattice gauge theories,
a theorem due to Tomboulis and Yaffe proves rigorously that the Wilson
loop has area law behavior if $\rho$ is non-zero 
:
\begin{equation}
\left\langle W\left(C\right)\right\rangle \le2\left[\frac{1}{2}\left(1-e^{-F_{\Lambda}}\right)\right]^{A/L_{1}L_{2}}\end{equation}
where $A$ is the area associated with $C$ \cite{Tomboulis:1985ah}.

%\subsection*{Issues}

There remain some unresolved questions with this construction. There
has been some criticism of both the implicit nature of the renormalization
group construction and its possible inability to distinguish between
asymptotically-free theories like $SU(2)$ and the non-asymptotically
free $U(1)$ lattice gauge theory \cite{Ito:2007nw,Tomboulis:2007mv,Ito:2008qu}.
It is true that the renormalization
group flow is defined implicitly, and there is not yet a complete
demonstration or working implementation of the specific renormalization
group procedure proposed. However, this is related to the second,
more physical criticism, which is based on the inability of Migdal-Kadanoff
renormalization group equations to differentiate between non-Abelian
and Abelian groups in the flow equations. This is in fact a very subtle
issue. The four-dimensional $U(1)$ lattice gauge theory has two phases:
a weak-coupling phase which is a lattice version of a free photon
theory, and a strong-coupling phase where magnetic monopoles are responsible
for confinement. After decades of effort, lattice simulations have
not yet conclusively determined the order of the transition between
these phases. It is believed that the weak-coupling phase consists
of a line of renormalization group fixed points, similar to that found
in the $d=2$ XY model, but no reliable renormalization group calculation
shows this. The Migdal-Kadanoff renormalization group shows a dramatic
change in the renormalization group flow near the known critical point:
the renormalization group beta function becomes very small, becoming
proportional to $\exp\left(-c/g^{2}\right)$. This behavior suggests
that the renormalization group is trying to account for magnetic monopole
effects. Exactly the same behavior is seen when the Migdal-Kadanoff
scheme is applied to the $d=2$ XY model \cite{Jose:1977gm}. 
In this case, there is a
reliable renormalization group calculation available, but it depends
on the introduction of a vortex fugacity and a renormalization group
flow with the coupling and the vortex fugacity as independent couplings.
In the generalized Migdal-Kadanoff renormalization group
developed by Tomboulis, 
the domain of decimation parameters is enlarged relative to
the original Migdal-Kadaonff scheme. The issue of the existence of a fixed point versus
a quasi-fixed point which fails by exponentially small corrections to be
a true fixed point in the U(1) gauge theory thus requires renewed
investigation in this more general context before it is settled.
Thus the original debate points to the large, difficult issue of accounting
for the infrared effects of topological excitations, particularly
within a renormalization group framework.

%\subsection*{Extension to Fermions}

Beyond issues concerning renormalization group flows in pure gauge
theories, there are important issues arising from the inclusion of
fermions within a real-space renormalization group framework. Fermions
appear in functional integrals as Grassmann variables, \emph{i.e.}
anticommuting $c$-numbers, and cannot be simulated or approximated
in the same way as bosonic variables. Fermion fields that appear only
quadratically in the action can be formally integrated over, yielding
a functional determinant that acts as an additional weight in the
bosonic functional integral. In lattice simulations, this functional
determinant is included, essentially as an additional, non-local interaction
that is expensive to simulate. In the context of real-space renormalization
group calculations, any partial integration over fermions with light
masses would introduce non-local interactions difficult to incorporate
into a renormalization group calculation. Furthermore, the bounds
on partition functions that hold for bosonic functional integrals
no longer hold for fermionic degrees of freedom. Nevertheless, there
has been some progress. Some analytical work has been carried out
for free fermions \cite{Balaban:1989ra,Wiese:1993cb},
 and there has been more recent
work in connection with the rooting problem 
of staggered lattice fermions \cite{Shamir:2006nj}. 
In work reported at the workshop,
Tomboulis discussed
a blocking technique for
fermions that is capable of producing a non-trivial fixed point in
$SU(2)$ lattice gauge theory for a sufficiently large number of flavors
$N_{f}$. The existence of such a fixed point is indicated by perturbation
theory, but the fixed point is generally located outside the region
of validity of perturbation theory. The presence of a non-trivial
infrared fixed point indicates the existence of a {}``conformal window''
which is of great interest for studies of particle physics beyond
the standard model \cite{Appelquist:1996dq,Appelquist:1998rb}.

\section{Confinement via the functional renormalization group and Dyson-Schwinger
equations}

The closely related functional renormalization group and Dyson-Schwinger
approaches to quark confinement have much in common with the research
on confinement on $R^{3}\times S^{1}$ described in section 3, including
the construction of the Polyakov loop effective potential. However,
these approaches focus on the infrared properties of conventional
gauge and ghost propagators. In the functional renormalization grouo
approach the renormalization group is used directly in attempting
to determine the non-perturbative infrared properties of gauge theories
from the renormalization group flow equation \cite{Braun:2007bx}. 
The Schwinger-Dyson
approach \cite{Alkofer:2008jy,Alkofer:2008tt}
aims at the calculation of the same quantities by a self-consistent
solution of the Schwinger-Dyson equations, which are the equations
of motion for correlation functions. 
Although beyond the scope of
this review, both approaches can be used to study the interplay of
confinement and chiral symmetry breaking \cite{Alkofer:2008tt,Braun:2009gm}, with results similar
to those obtained from more phenomenological approaches such as the
PNJL model \cite{Fukushima:2003fw}.

%\subsection*{functional RG}

The functional renormalization group can be elegantly described for
any field theory by the RG flow equation for the effective action
$\Gamma\left[\phi\right]$ \cite{Berges:2000ew,Pawlowski:2005xe}. 
The effective action $\Gamma\left[\phi\right]$
is the generator of one-particle irreducible (1PI) vertices, where
$\phi$ is shorthand notation for one or more classical fields, one
for each quantum field. The $n$-point 1PI vertices are obtained from $\Gamma\left[\phi\right]$
by functional differentiation:\begin{equation}
\Gamma^{(n)}\left(x_{1},...,x_{n}\right)=\frac{\delta^{n}\Gamma}{\delta\phi\left(x_{1}\right)...\delta\phi\left(x_{n}\right)}\end{equation}
The $2$-point function $\Gamma^{(2)}(x_{1},x_{2})$ is the full propagator.
The functional renormalization group generalizes $\Gamma\left[\phi\right]$
to $\Gamma_{k}\left[\phi\right]$ , which is identical to $\Gamma\left[\phi\right]$,
except that only momenta with $q^{2}\gtrsim k^{2}$ are included in
the $n$-point vertices. This exclusion of infrared modes is accomplished
by inserting an infrared cutoff function $R_{k}$ into the functional integral
that defines $\Gamma\left[\phi\right]$, so that by definition $\Gamma_{0}\left[\phi\right]=\Gamma\left[\phi\right]$.
The RG equation for $\Gamma_{k}\left[\phi\right]$ is\begin{equation}
\frac{\partial}{\partial k}\Gamma_{k}\left[\phi\right]=\frac{1}{2}Tr\left\{ \frac{1}{\Gamma_{k}^{(2)}\left[\phi\right]+R_{k}}\frac{\partial}{\partial k}R_{k}\right\} \end{equation}
 where the trace involves an integration over momenta or coordinates
as well as summation over any internal indices. For the study of spontaneous
symmetry breaking, the key quantity is typically the effective potential
$V\left(\phi\right)$, obtained by setting $\phi$ to a space-time
independent value, and dividing $\Gamma\left[\phi\right]$ by the
volume $\Omega$ of space-time: $V\left(\phi\right)=\Gamma\left[\phi\right]/\Omega$.

%\subsection*{application to gauge theories}

The functional renormalization group formalism can be applied to gauge
theories with some modifications of the procedure for scalar fields:
a gauge-fixing term must be added to the action and ghost fields must
be introduced. In an asymptotically free theory, one can approximate
$\Gamma_{k}$ by the classical action $S$ for large $k$, and evolve
the functional RG equation down to $k=0$. The effective potential
for $A_{4}$ is given by $V\left(A_{4}\right)=\Gamma\left[A_{4}\right]/\Omega$,
where $A_{4}$ is a constant, representing a constant Polyakov loop.
This effective action can be evaluated using the background field method,
and
the effective potential is obtained within the functional
renormalization group approach
from\begin{equation}
V_{k}\left(A_{4}\right)=\frac{\Gamma_{k}\left[A_{4}\right]}{\Omega}\end{equation}
 in the limit $k\rightarrow0$. 
 In principal, the functional renormalization
group is exact, but in practice, approximations are necessary. A scaling
solutions is assumed in the infrared, in which the transverse gauge
field propagator scales as $1/\left(p^{2}\right)^{1+\kappa_{A}}$
and the ghost propagator scales as $1/\left(p^{2}\right)^{1+\kappa_{C}}$.
Then at sufficiently low temperature, the effective potential is given
by\begin{equation}
V\left(A_{4}\right)=\left\{ \frac{d-1}{2}\left(1+\kappa_{A}\right)+\frac{1}{2}-\left(1+\kappa_{C}\right)\right\} \cdot\frac{1}{\Omega}Tr\log\left[-D^{2}\left[A_{4}\right]\right]\end{equation}
 where $-D^{2}\left[A_{4}\right]$ is the standard scalar propagator
for a particle at finite temperature in a background field $A_{4}$.
This should be compared with the standard high-temperature expression\begin{equation}
V\left(A_{4}\right)=\left\{ \frac{d-1}{2}+\frac{1}{2}-1\right\} \cdot\frac{1}{\Omega}Tr\log\left[-D^{2}\left[A_{4}\right]\right].\end{equation}
 If the solution of the functional renormalization group equations
satisfies\begin{equation}
\frac{d-1}{2}\left(1+\kappa_{A}\right)+\frac{1}{2}-\left(1+\kappa_{C}\right)<0\end{equation}
 then the sign of the effective potential changes between high- and
low-temperature, leading to the same confining minimum at low temperature
found at high temperature in section 3. 
This scenario was studied in Braun, Gies, and Pawlowski 
\cite{Braun:2007bx}
using Landau-DeWitt
gauge, and the values $\kappa_{A}=-1.19...$ and $\kappa_{C}=0.595...$
were obtained, indicating confinement.

%\subsection*{Relation to Kai}

A closely related approach is based on the self-consistent solution
of the Dyson-Schwinger equations in the infrared \cite{Alkofer:2008jy}. 
The Dyson-Schwinger
equations are an infinite set of coupled equations relating $n$-point
correlation functions to correlation functions of higher order. Again
using a shorthand notation, the Dyson-Schwinger equations may be derived
from functional integration by parts\begin{equation}
\int\left[d\phi\right]\frac{\delta}{\delta\phi\left(y\right)}\left[\phi\left(x_{1}\right)...\phi\left(x_{n}\right)e^{-S\left[\phi\right]}\right]=0\end{equation}
The simplest case of $n=0$ reduces to $\left\langle \delta S/\delta\phi\left(y\right)\right\rangle =0$,
the statement that the classical equation of motion holds on the average.
In both the functional renormalization group and Dyson-Schwinger approaches,
an infinite tower of coupled functional equations for the correlation
functions are solved, and both approaches include
the same physics, albeit captured
in different ways.
At present, studies using the Schwinger-Dyson approach have
restricted themselves mainly to the infrared regime
relying on a general power counting analysis of the possible IR fixed points of the full system of equations of quenched QCD, {\it i.e.}, without quarks, in the Landau gauge. 
Applying this approach
to gauge theories requires extensive graphical analysis to identify
the most important contributions to the Dyson-Schwinger equations
in the infrared. Choice of gauge is important in simplifying the analysis.
One finds both a decoupling solution that does not feature 
any infrared enhancement and a scaling solution where infrared 
divergences are induced in the ghost sector \cite{Alkofer:2008jy}. 
In the scaling solution, a long range interaction in gauge-dependent
local correlation functions in the quark sector arises from a strong kinematic infrared divergence of the quark-gluon vertex. 
In this approach, the Wilson loop is then given by 
an infinite series of these local Green functions and
the infrared divergence of the leading term, given by the quark-quark scattering kernel, 
yields in the quenched approximation area law behavior \cite{Alkofer:2008tt}.
Recently work indicates that the 
corresponding long range interaction is a 
universal property of the gauge sector 
\cite{Fister:2010yw}.
Although there are some significant differences
at a technical level, the Dyson-Schwinger
and functional renormalization group approaches are very similar,
and in Landau gauge, the infrared behavior
found for the ghost and gluon propagators is essentially the same
\cite{Alkofer:2008jy,Fischer:2008uz}.

\section{Conclusions}

The problem of quark confinement is an important and vital subject
in theoretical physics, connected not just to hadronic physics, but
to many areas of physics by concepts and techniques. It is the quintessential
problem of dimensional transmutation, where renormalization introduces
a length scale into a classicaly scale-invariant system. Fundamental
questions of mechanism, \emph{i.e., }the cause of confinement, have
not yet been answered. In addition to questions of mechanism, there
are also questions surrounding the implications of confinements, \emph{e.g.},
string tension scaling laws and other string properties.

Approaches have been developed in the last few years that are shedding
new light on these problems. The discovery of a rich set of new phases
in gauge theories on $R^{3}\times S^{1}$ shows that there are new
phenomena that remain to be uncovered. The topics discussed here,
despite differences of approach and emphasis, have more in common
with each other than the use of the renormalization group. For example,
recent work using the functional renormalization group and the Dyson-Schwinger
equation shares with work on $R^{3}\times S^{1}$ an emphasis on the
effective potential for the Polyakov model. As discussed in section
3, there are issues concerning magnetic monopoles raised by recent
work using real-space renormalization group methods that are related
to recent work on $R^{3}\times S^{1}$. Beyond common technical issues,
there are two basic themes all the approaches discussed here have
in common: an incorporation of both the physics of continuum gauge
theories indicated by the renormalization group and the increasingly
precise results of lattice gauge theory. This marks a significant
departure from previous approaches to confinement, which tended to neglect either
lattice or continuum results, and should be welcomed as genuine progress
towards greater understanding.

\begin{acknowledgments}
I would like to thank
Mike Birse, Yannick Meurice and Shan-Wen Tsai 
for organizing the workshop and for asking me
to write this review.
I am also deeply grateful to my colleagues
Hiro Nishimura, Jan Pawlowski, Kai Schwenzer and Terry Tomboulis,
for their help in preparing this review.
This work was supported by the U.S. Dept. of Energy.
\end{acknowledgments}

\end{document}